# Laser-Induced Heating in Diamonds: Influence of Substrate Thermal Conductivity and Interfacial Polymer Layers

*Md Shakhawath Hossain,[1] Jiatong Xu,[1] Thi Ngoc Anh Mai,[1] Nhat Minh Nguyen,[1] Trung Vuong Doan,[1] Chaohao Chen,[2] Qian Peter Su,[3] Yongliang Chen,[4] Evgeny Ekimov,[5] Toan Dinh,[6,7] Xiaoxue Xu,[3] and Toan Trong Tran [1, \*]*

[1]School of Electrical and Data Engineering, University of Technology Sydney, Ultimo, NSW, 2007, Australia.

[2]School of Mathematical and Physical Sciences, Faculty of Science, University of Technology Sydney, NSW 2007, Australia

[3]School of Biomedical Engineering, University of Technology Sydney, Ultimo, NSW, 2007, Australia.

[4]Department of Physics, The University of Hong Kong, Pokfulam, Hong Kong, China.

[5]Vereshchagin Institute for High Pressure Physics, Russian Academy of Sciences, Troitsk 142190, Russia.

[6]School of Engineering, University of Southern Queensland, Toowoomba, Queensland 4350, Australia

[7]Center for Future Materials, University of Southern Queensland, Toowoomba, Queensland 4350, Australia

\*Corresponding author: trongtoan.tran@uts.edu.au

**Abstract**

Diamonds hosting color centers possess intrinsically high thermal conductivity; therefore, laser-induced heating has often received little attention. However, when placed on substrates with low thermal conductivity, localized heating of diamonds under laser excitation can become significant, and the presence of an interfacial polymer layer between substrate and diamond further amplifies this effect. Yet, the relationship between substrate thermal conductivity, polymer thickness, and laser heating remains to be established. Here, a systematic investigation is presented on laser-induced heating of silicon-vacancy diamond on substrates with varying thermal conductivity and interfacial polymer thickness. Results reveal that even at a low excitation power of 737 µW/µm$^2$, thin amorphous holey carbon—the lowest-conductivity substrate (~0.2 W m$^{-1}$ K$^{-1}$) studied—exhibits substantial heating, while glass (~1.4 W m$^{-1}$ K$^{-1}$) and polydimethylsiloxane (PDMS) (~0.35 W m$^{-1}$ K$^{-1}$) show noticeable heating only above 2.95 mW/µm². For polymer interlayers, a thickness of just 2.2 µm induces significant heating at 2.95 mW/µm² and above, highlighting strong influence of both substrate and polymer thickness on local heating response. Experimental findings are further validated using COMSOL Multiphysics simulations with a steady-state 3D heat transfer model. These results provide practical guidance for substrate selection and sample preparation, enabling optimization of conditions for optical thermometry and quantum sensing applications.

**1. Introduction**

Nanoscale optical thermometry[1-4], also known as luminescence thermometry, enables precise measurement of temperature variations at the nanometer scale. Over the past decade, it has rapidly emerged as a multifaceted and impactful research frontier with broad implications for fields ranging from advanced materials science to biomedicine[5]. Breakthroughs in materials science alongside the continuous evolution of nanotechnologies—including nanoelectronics[6], nanophotonics[7] and cell imaging[8] have significantly driven progress in this research area. The key advantage of optical thermometry lies in its ability to detect temperature remotely. Although scanning thermal microscopy (SThM)[9-11], a widely used thermometry technique, provides high spatial resolution, it suffers from complex heat-transfer challenges that may introduce large uncertainties through thermalization of the sample within the scanning device. In contrast, optical thermometry overcomes this limitation, enabling remote and minimally invasive nanoscale temperature measurements with a more straightforward calibration process for improved accuracy.

Theoretically, optical thermometry relies on external optical excitation to induce luminescence from a specific phosphor[12], such as color centers in nanodiamonds[13-20], quantum dots[21], organic dyes[22], or upconversion nanoparticles[23]. Phosphor emission properties—such as signal intensity, line shape, emission wavelength, polarization state, excited-state lifetime, and full width at half maximum (FWHM)—vary with temperature.[24] Quantifying these dependencies enables mapping of the local temperature of the underlying surface in contact with the phosphor. This method falls into the category of semi-contact thermometry, as the sensor probe is directly placed on the sample while the optical emission is collected in the far field.

Among various phosphors, color centers (optically active defects point) in the diamond lattice stand out as promising thermal sensors owing to diamond's small size, wide bandgap, ultra-high thermal conductivity, exceptional photophysical properties, mechanical and chemical stability[25, 26], and compatibility with both delicate biological applications and harsh environments. Several all-optical thermometry techniques have already been reported,[19, 27, 28] leveraging the properties of color centers embedded within the diamond crystal lattice. However, an often-overlooked aspect of all-optical thermometry is laser-induced self-heating: diamonds with impurities and surface graphitic layers absorb part of the incident light and dissipate it non-radiatively, resulting in optical heating that can affect the spectral properties of the emission due to the elevated temperature. The extent of laser-induced heating depends on how efficiently heat can dissipate through the diamond–substrate interface, and is exacerbated by substrates with low thermal conductivity[29] or by residual polymer interlayers between the thermal sensor and the substrate, which hinder effective thermal transport. As a result, the same excitation power can yield different measured temperatures, governed by the substrate's thermal conductivity and the cleanliness of the diamond–substrate interface. Acknowledging this phenomenon is essential for obtaining reliable measurements when deploying pre-calibrated, transferable thermal sensors on platforms that differ from the calibration substrate in terms of thermal conductivity or surface cleanliness. Undetected interlayers or substrates with low thermal conductivity can induce significant optical heating, leading to discrepancies between the measured and actual temperatures. Such deviations contribute to increased thermal-equivalent noise (TEN)[30], thereby compromising measurement accuracy. To the best of our knowledge, a systematic study quantifying how substrate thermal conductivity and interfacial polymer residues influence laser-induced heating in color-center diamonds has not yet been reported.

To address this gap, we present a comprehensive study of laser-induced heating in silicon-vacancy (SiV⁻) diamonds, focusing on the roles of underlying substrate thermal conductivity and interfacial polymer layer thickness. To further elucidate the underlying thermal response, we employed finite element method (FEM) simulations in COMSOL Multiphysics software to model the spatial temperature distribution of SiV⁻ diamonds on the experimentally investigated substrates. The motivation of this work is thus threefold: first, to quantify the laser power required to induce significant heating on substrates of varying thermal conductivities; second, to determine the minimum polymer layer thickness at which interfacial effects begin to produce measurable heating; and third, to use simulations to validate the accuracy of our experimental findings. Thus, this work establishes design principles for minimizing thermal artifacts in diamond-based nanoscale thermometry and quantum devices.

## 2. Results and Discussion

In this work, the SiV$^-$ diamond was employed as the active temperature-sensing element due to its easily detectable narrow zero-phonon line (ZPL) bandwidth at room temperature and the higher concentration of emission power in its photoluminescence (PL) spectrum. To establish a temperature-dependent spectral reference, diamonds containing a high concentration of SiV$^-$ color centers (~1.9×10$^{18}$ cm$^{-3}$) were used. These diamonds were synthesized via the high-pressure high-temperature (HPHT) method **(cf. 4.1 Sample Preparation)**. They were calibrated on a bulk diamond substrate, chosen for its exceptionally high thermal conductivity (~2400 W m$^{-1}$ K$^{-1}$).[31] The diamond substrate has a high thermal conductivity, which helps to minimize laser-induced heating. It rapidly dissipates heat through the diamond–substrate interface. As a result, the observed spectral shifts represent the actual external temperature rather than photoinduced thermal effects. Thus, the bulk diamond substrate serves as a reliable thermal reference for calibrating temperature-dependent photoluminescence spectrum.

For calibration, diamonds were first drop-cast onto a pre-cleaned bulk diamond substrate and annealed at 200 °C for 20 minutes to remove surface residues and enhance particle–substrate adhesion. **Figure S1 (cf. Supporting Information)** presents the Raman spectrum of a representative microdiamond after annealing at 200 °C. The sharp and symmetric first-order Raman mode observed at ~1335 cm$^{-1}$ confirms the presence of crystalline *sp³*-bonded carbon, indicative of the diamond phase and high crystal quality. Furthermore, the absence of a G-band around 1580 cm$^{-1}$ indicates that no graphitic or amorphous carbon is present on the microdiamond surface. The bulk diamond substrate was then glued on a high-precision temperature controller (Microoptik MHCS600) using silver conductive paste to improve thermal coupling between the substrate and the controller. The temperature gradually increased from 25 °C to 85 °C in 10 °C increments using the temperature controller. At each temperature point, photoluminescence spectra of the SiV$^-$ diamond were captured using a custom-built confocal microscopy setup. The PL spectrum corresponding to each controller-measured temperature set point was considered the 'true' value for subsequent analysis. **Figure S2a (cf. Supporting Information)** shows the PL spectra of a representative diamond recorded at various temperatures, revealing a clear red shift of the zero-phonon line, along with linewidth broadening and a decrease in PL intensity as the temperature increases. This redshift is primarily attributed to enhanced electron–phonon coupling at elevated temperatures, which leads to a reduction in the ZPL energy and thus shifts the emission to longer wavelengths.[32] Concurrently, at higher temperature, phonon-induced dephasing broadens the linewidth, while phonon interactions promote non-radiative relaxation, reducing radiative recombination and thus decreasing the SiV$^-$ emission intensity.[33] For this study, the ZPL value was tracked as the spectral feature representing temperature, as it exhibits a significant linear red shift with increasing temperature. **Figure S2b (cf. Supporting Information)** illustrates the temperature-dependent shift of the ZPL, fitted with a linear function to serve as the calibration curve. This calibration was performed using five microdiamonds of comparable size and morphology, with error bars representing the standard deviation of the ZPL position at each temperature. The data points show excellent agreement with the linear fit, yielding an $R^2$ value of approximately 0.99, and the small error bars further support the consistency and reliability of the calibration.

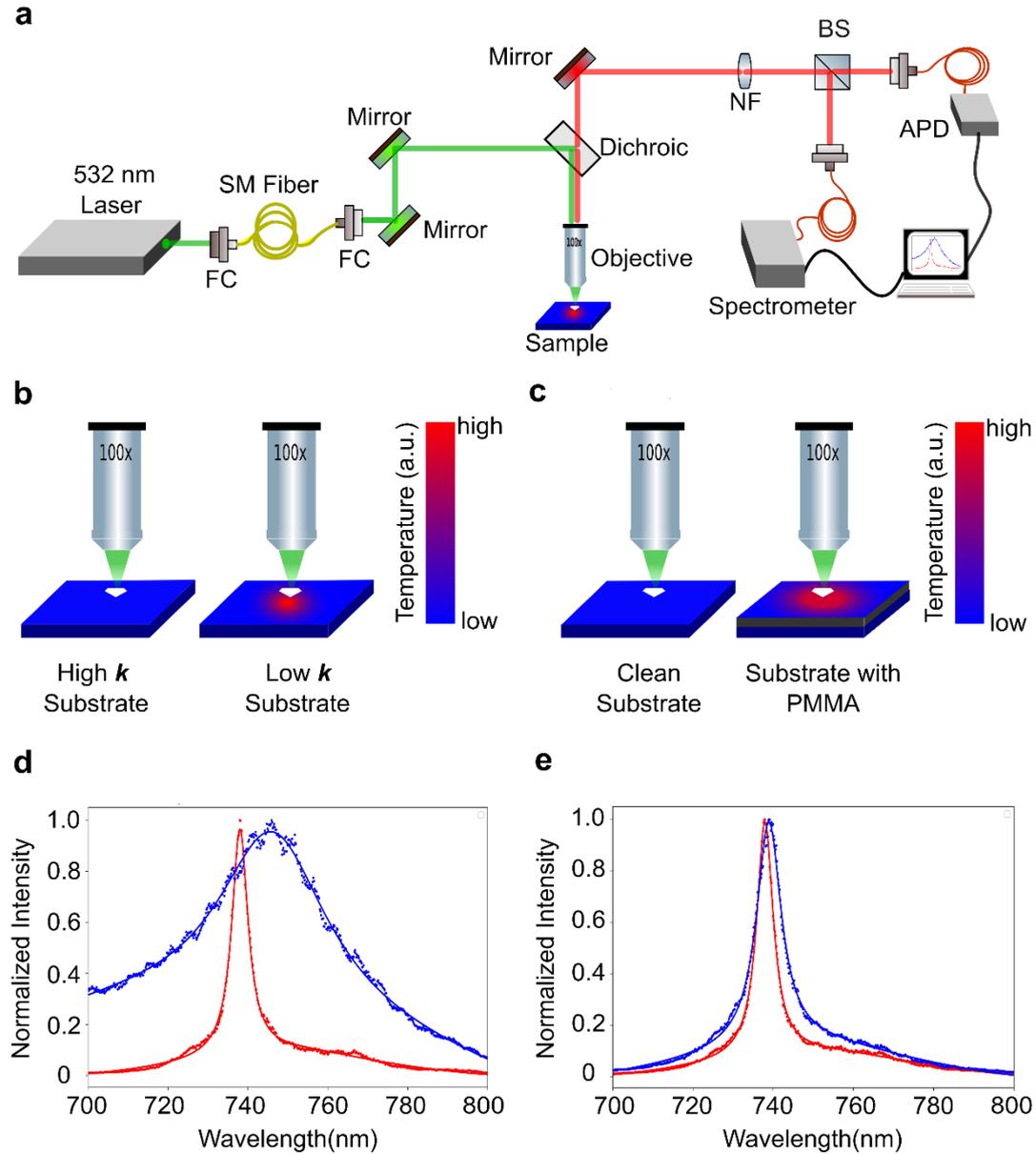

**Figure 1.** Overview of the experimental setup and the conceptual illustration of laser-induced heating on different substrates studied. **(a)** Schematic of the experimental setup. Green and red lines represent the excitation beam and the fluorescence beam, respectively. The dichroic is used to reflect the 532 nm excitation laser to the sample and transmit the photoluminescence emission signal emitted from SiV⁻ diamond. A 532-nm notch filter further filters the emission, which is then collected by a spectrometer and an APD. Abbreviations: APD (avalanche photodiode); FC (fiber coupler); SM (single mode); BS (beam splitter); NF (notch filter). **(b)** Conceptual illustration of laser-induced heating in diamonds, comparing temperature distributions on low and high-thermal-conductivity ($k$) substrates. The same excitation power produces a higher heating on low-$k$ substrates. **(c)** Schematic illustrating how an interfacial layer alters heat flow: a microdiamond on a clean substrate (left) versus on a PMMA layer (right), with the latter exhibiting a higher temperature profile for the same excitation power. **(d)** PL spectra of a SiV⁻ microdiamond, fitted with a double-Lorentzian function, under the same 8 mW

excitation on high and low thermal conductivity substrates—bulk diamond (red) and an amorphous holey carbon (blue). On the low-conductivity, thin, amorphous, holey carbon, the emission red-shifts and broadens, indicating greater laser-induced heating than on diamond. **(e)** PL spectra of a SiV$^-$ diamond, fitted with a double-Lorentzian function, under the same 8 mW excitation on a clean substrate (red) and on a substrate coated with the highest thickness of PMMA layer (16 µm) (blue). The interface PMMA layer also causes the emission to red-shift and broaden, indicating a higher temperature readout than on a clean substrate.

**Figure 1a** shows the schematic diagram of our custom-built confocal setup (**cf. 4.2 Optical Measurement**), where a continuous-wave 532 nm laser serves as the excitation source, focused on a spot diameter of about 0.93 µm through a high-NA (0.70) air objective. A 649-nm dichroic mirror separates the excitation and collection paths; since the SiV$^-$ emission is centered around ~739 nm, it is well within the transmission band of the dichroic. Additionally, a 532 nm notch filter is placed in the detection path to suppress residual laser noise before the signal reaches the APD and spectrometer. **Figures 1b** and **1c** illustrate the conceptual framework of this work, highlighting how laser-induced heating of the SiV$^-$ diamond is influenced by the thermal conductivity of the substrate and the presence of an interfacial layer between substrate and diamond. Hypothetically, when a 532 nm green laser is focused on a SiV$^-$ containing diamond, a portion of the optical energy is absorbed and converted into heat. The amount of absorption depends on several factors, including the concentration of SiV$^-$ centers, the quality of the diamond crystal (e.g., presence of defects or impurities), and the size of the diamond. This heat must dissipate into the underlying substrate, in accordance with the principles of diffusive heat transport.[34] The rate and extent of this thermal diffusion are governed by the thermal conductivity of the substrate material, which plays a crucial role in determining the local temperature rise of the diamond. If the substrate has low thermal conductivity (**low *k***), it cannot efficiently dissipate heat, leading to a temperature rise in the diamond. In contrast, a high thermal conductivity (**high *k***) substrate enables faster heat diffusion, resulting in negligible laser-induced heating of the diamond. On the other hand, even in the presence of a highly conductive substrate, an interfacial polymer layer can modify heat flow by introducing additional thermal resistance. Here, PMMA (polymethyl methacrylate) was employed to replicate the polymer-like residue commonly encountered after transfer or fabrication processes. This added resistance reduces the efficiency of heat transfer from the diamond to the substrate, resulting in an elevated local temperature within the diamond. The thickness of the PMMA layer also plays a key role in modulating this thermal decoupling effect, with thicker layers resulting in greater temperature rise. **Figures 1d and 1e** show the PL spectra of the SiV$^-$ diamond under different experimental conditions, measured at the same excitation power (11.8 mW/μm²). **Figure 1d** compares normalized PL spectra on substrates with the highest (bulk diamond) and lowest (thin amorphous holey carbon) thermal conductivity. In contrast, **Figure 1e** presents PL spectra on substrates with minimal and maximal PMMA residue. The implications of these results will be discussed in detail later.

In the first part of this experiment, we systematically studied the relationship between laser power and the induced heating of SiV⁻ diamonds deposited on five substrates with thermal conductivities ($k$) spanning from very high to very low: bulk diamond, silicon (Si), glass (SiO₂), PDMS, and thin amorphous holey carbon (TEM grid). Five bright diamonds with diameters of (5.0 ± 1.5) μm were measured on each substrate to ensure statistical reliability and represent the variability in the photoluminescence response. Throughout the experiment, the laser spot on the samples was kept constant, with an area of 0.68 μm², while the microdiamonds used were significantly larger than the spot, resulting in localized irradiation of only a portion of each diamond. The laser excitation power was initially set to 1 μW (1.48 μW/μm²) and then increased stepwise to a maximum of 8 mW (11.8 mW/μm²). A variable neutral density filter was used to adjust the laser power, which was measured at the back aperture of the objective.

**Figure 2a** presents the ZPL heat map of an SiV⁻ diamond on a bulk diamond substrate, with the x-axis representing wavelength, the y-axis laser power, and the color bar denoting PL intensity. **Figure 2b** shows the average ZPL position and the corresponding estimated temperature, obtained from the calibration curve, for five SiV⁻ microdiamonds on a bulk diamond substrate as a function of laser power. The absence of any significant ZPL shift, even at maximum laser power, indicates negligible or no laser-induced heating due to the high thermal conductivity of the bulk diamond substrate. Similarly, **Figures 2c** through **2j** present the ZPL heat maps and corresponding temperature versus laser power plots for SiV⁻ diamonds on other substrates, including silicon, glass, PDMS, and thin amorphous holey carbon, highlighting substrate-dependent thermal behavior. In the case of silicon, which has moderate thermal conductivity (~150 W m⁻¹ K⁻¹),[35] the absence of a significant ZPL shift with increasing laser power—almost similar to that observed for bulk diamond—indicates efficient heat dissipation. However, glass (~1.4 W m⁻¹ K⁻¹)[36] and PDMS (~0.35 W m⁻¹ K⁻¹) [37], both of which are significantly less thermally conductive than bulk diamond and silicon, exhibit pronounced ZPL shifts when the laser power exceeds 2 mW, suggesting significant laser-induced heating. At the highest laser power of 8 mW (11.8 mW/μm²), the temperature rises to approximately 50 °C for glass and 95 °C for PDMS, as shown in **Figure 2f** and **2h**. Remarkably, the thin amorphous holey carbon substrate exhibits a significant ZPL shift even at 500 μW (737 μW/μm$^2$), and the temperature rises nearly ~560 °C at 8 mW, reflecting significant local heating caused by its low thermal conductivity (~0.2 W m$^{-1}$ K$^{-1}$)[38], as illustrated in **Figures 2i** and **2j**. At 8 mW laser power, the temperature rise is ~2 times greater in glass, ~3.8 times greater in PDMS, and nearly ~10.5 times higher in thin amorphous holey carbon, compared to bulk diamond. **Figure 1d** shows the spectra acquired from a bulk diamond and a thin amorphous holey carbon sample under the same excitation power of 8 mW. A main peak is observed at ~738 nm for the bulk diamond substrate and ~745 nm for the thin amorphous holey carbon substrate, corresponding to the luminescence of the SiV$^-$ center. Interestingly, the ZPL wavelength, FWHM, and corresponding temperature of SiV$^-$ centers observed on the thin amorphous holey carbon substrate at 8 mW laser power are in excellent agreement with the values reported by Lagomarsino et al.[39] Luminescence from the thin amorphous holey carbon exhibits significant broadening, with a full width at half maximum (FWHM) of ~20 nm compared to ~5 nm for the bulk diamond, indicating substantial local heating. **Figure S3a–e (cf. Supporting Information)** shows the measured temperatures of five different SiV$^-$ diamonds on five different substrates. Some temperature variations are observed between microdiamonds on the same substrate, which can be attributed to differences in structural defect concentration and distribution, particle size, shape, orientation, and local thermal contact with the substrate — all factors that can influence heat dissipation efficiency. These temperature deviations become more pronounced at higher laser powers, particularly on substrates with low thermal conductivity, due to instability and increased uncertainty at elevated temperatures.

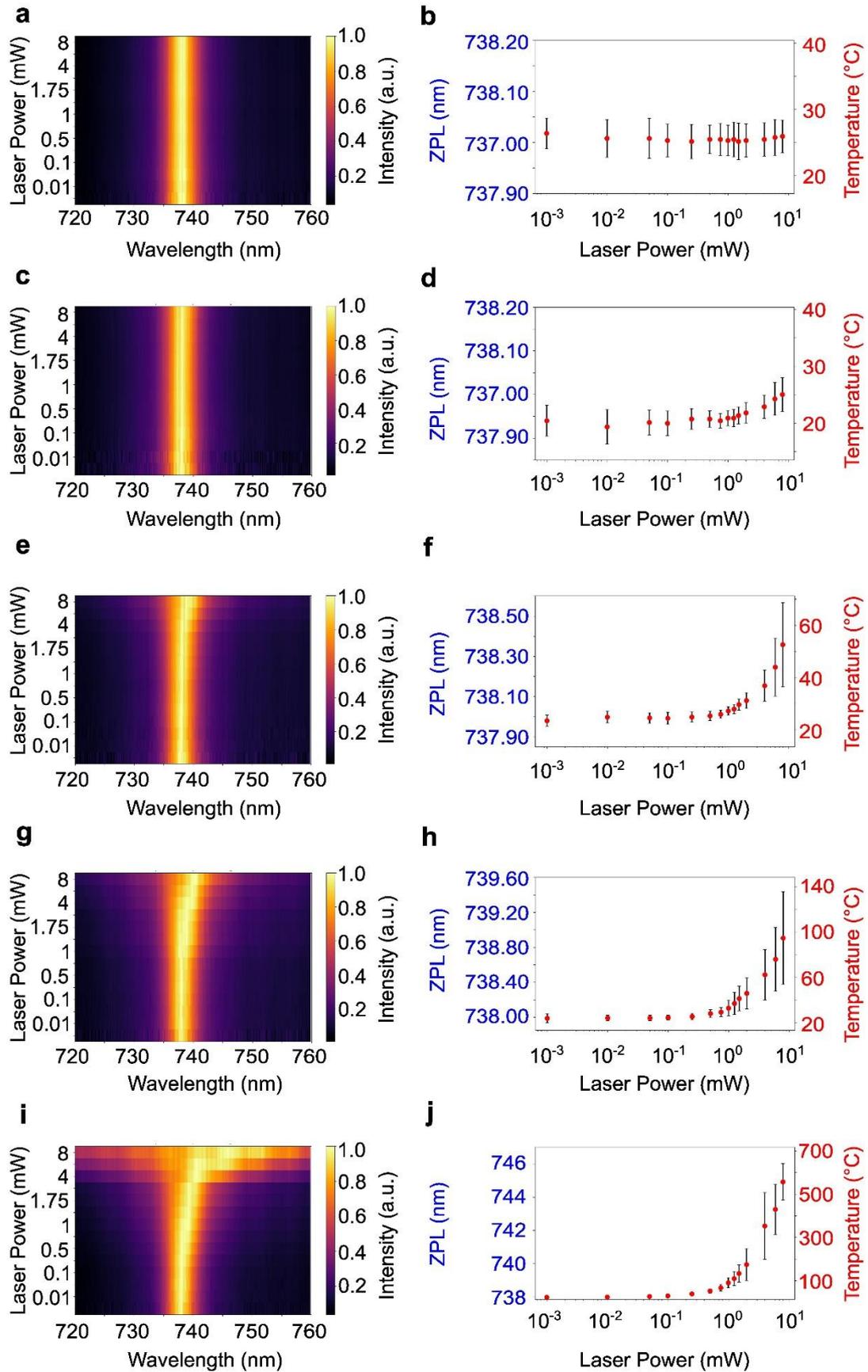

**Figure 2.** Laser-induced heating response of SiV⁻ diamonds on substrates with different thermal conductivities. **(a, b)** Bulk diamond, **(c, d)** silicon, **(e, f)** glass, **(g, h)** PDMS, and **(i, j)** thin amorphous

holey carbon. Heat maps **(a, c, e, g, i)** show the ZPL spectral region of a representative SiV⁻ diamond, with photoluminescence (PL) intensity plotted as a function of laser excitation power (y-axis) and wavelength (x-axis); color denotes PL intensity. Corresponding plots **(b, d, f, h, j)** display the average ZPL peak position and the estimated temperature as a function of laser power for five different diamonds on each substrate, calculated from the calibration curve. The error bars reflect the variability (standard deviation) across measurements from five distinct microdiamonds at each laser power. The x-axis of the right panels is presented on a logarithmic scale.

The second part of the project focused on investigating how the thickness of the interfacial residue between the thermal sensor and the substrate influences laser-induced heating. This is particularly important because, in many micro- and nanofabrication workflows, thin polymer films or residual photoresist layers often remain at the thermal sensor–substrate junction, introducing additional thermal boundary resistance of poorly understood magnitude. For this part of the experiment, silicon was chosen as the base substrate since it exhibits negligible laser-induced heating, as verified in our first part of the experiment. Five silicon-based samples were chosen: a clean silicon substrate without PMMA residue, and substrates coated with PMMA layers of 590 nm, 2.2 µm, 5.2 µm, and 16 µm thickness. Different PMMA thicknesses were prepared using a dip-coating process by varying the withdrawal speed **(cf. 4.1 Sample Preparation). Figure S4 (cf. Supporting Information)** presents profilometer measurements of PMMA layers with varying thicknesses. Consistent with the methodology outlined in the first part of this study, five diamonds were measured on each substrate. Laser excitation power was incrementally increased from an initial value of 100 µW (147.5 µW/µm$^2$) to a maximum of 8 mW (11.8 mW/µm$^2$). **Figures 3a–3j** present the ZPL heat maps and the corresponding temperature versus laser power plots for SiV⁻ microdiamonds on five different substrates: a bare silicon substrate (no PMMA), and silicon substrates coated with PMMA layers of 590 nm, 2.2 µm, 5.2 µm, and 16 µm thickness. **Figures 3b** and **3d** show that no significant laser-induced heating is observed for substrates without PMMA or even with a 590 nm PMMA layer. However, for PMMA thicknesses above 2.2 µm, laser heating becomes evident beyond 2 mW (2.95 mW/µm$^2$) excitation power. As the PMMA thickness increases, the extent of laser-induced heating also increases, as illustrated in **Figures 3f, 3h,** and **3j**. This can be clearly understood as thicker PMMA reduces heat dissipation to the underlying high-conductivity substrate, resulting in a higher local temperature in the microdiamond under the same laser power. Notably, for the thickest PMMA layer (16 µm), heating is observed even with 500 µW (737 µW/µm$^2$) laser power, with a temperature increase of approximately 85 °C at 8 mW laser power. **Figure S5a–e (cf. Supporting Information)** shows the measured temperatures of five different SiV⁻ diamonds on substrates described above. Consistent with the first part of the experiment, some inter-diamond temperature variations were observed. This variability likely arises from nonuniform PMMA thickness, differences in diamond defect concentration, size, and variations in local thermal contact.

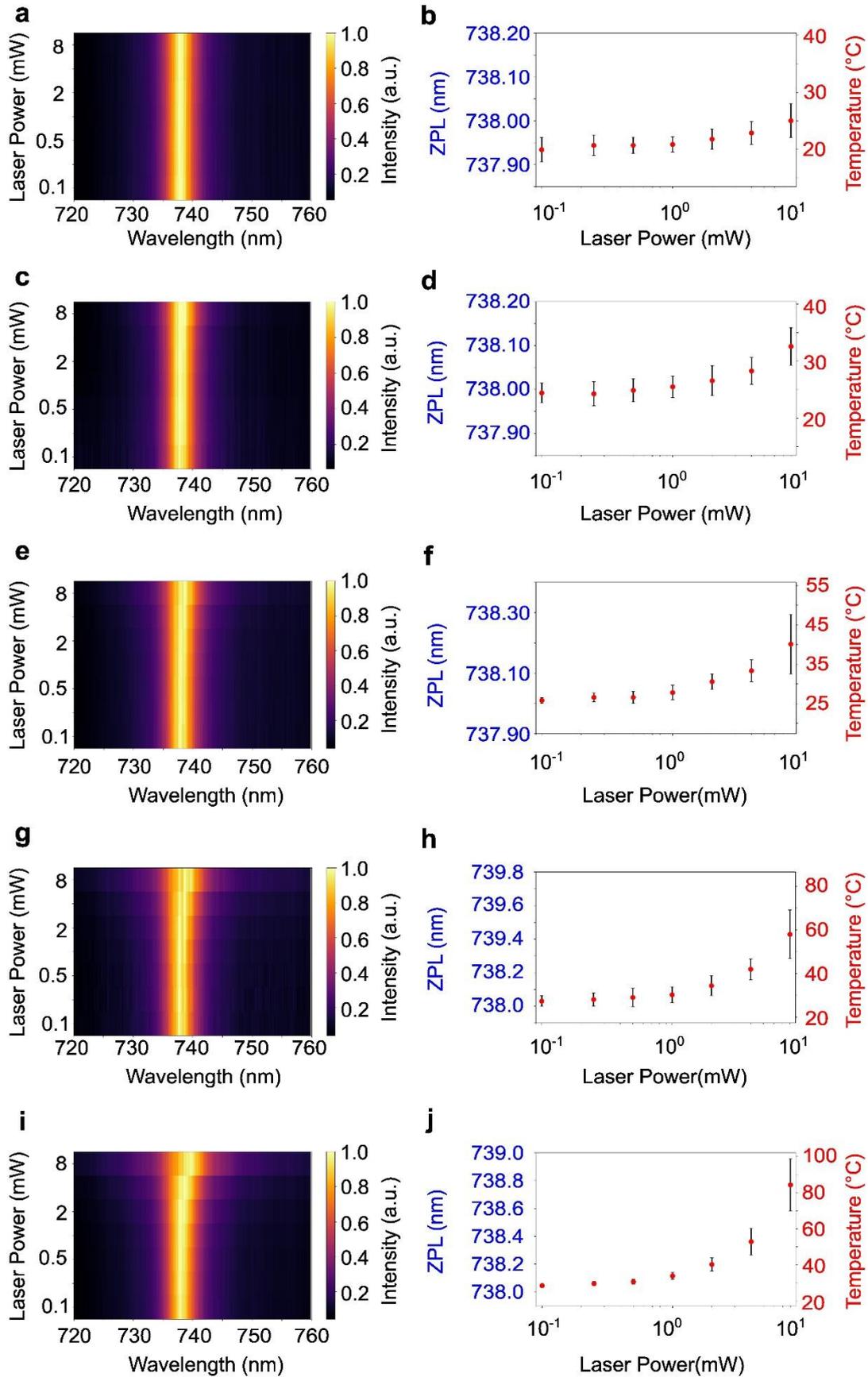

**Figure 3.** Laser-induced heating response of SiV⁻ diamonds on substrates coated with PMMA layers of different thicknesses. Panels **(a, b)** correspond to substrates without PMMA; **(c, d)** to 590 nm PMMA;

**(e, f)** to 2.2 μm PMMA; **(g, h)** to 5.2 μm PMMA; and **(i, j)** to 16 μm PMMA. Heat maps in **(a, c, e, g, i)** show PL intensity of the ZPL region for a representative SiV$^-$ microdiamond as a function of laser power and wavelength. Corresponding plots in **(b, d, f, h, j)** present the average ZPL peak position and extracted temperature versus laser power for five microdiamonds per substrate, using the same calibrated temperature–wavelength relationship. The error bars indicate the standard deviation of measurements taken from five different microdiamonds at each point. The x-axes in the right panels are displayed on a logarithmic scale.

To better understand and validate the accuracy of our thermal measurements, we performed finite element method (FEM) simulations to predict the temperature rise in the three-dimensional heating geometry used experimentally. Specifically, 3D heat distribution simulations were carried out in COMSOL Multiphysics (v.5.5), replicating the experimental conditions for SiV$^-$ diamonds on substrates of different thermal conductivity and with different PMMA layer thicknesses. A focused 532 nm Gaussian laser beam was modeled as a heat source applied to the top surface of an SiV$^-$ microdiamond placed on a different substrate (**cf. 4.3 Comsol Simulation).**

In the first part of the simulation, the steady-state temperature distribution of the diamond was evaluated on substrates with different thermal conductivities under 8 mW laser excitation, which was the highest laser power used in the experiment. The laser spot area (~0.68 μm$^2$) is much smaller than the top surface of the microdiamond (~10.8 μm$^2$ for a 5 μm edge-length octahedron), so only a small portion of the diamond is illuminated relative to its overall size. The substrates were modeled with dimensions of 200 μm × 200 μm to ensure uniformity. **Figure 4 (a-e)** presents the simulation results for diamonds on substrates with different thermal conductivities, highlighting the influence of substrate conductivity on the steady-state temperature distribution. In high thermal conductivity substrates, heat was efficiently dissipated from the laser focal point, resulting in negligible temperature rise. In contrast, low thermal conductivity substrates retained heat near the laser focus, leading to pronounced localized heating. These findings agree with the experimental PL shifts and ZPL heating trends observed in the earlier sections of this work. **Figure 4(f)** shows a comparison between simulated and experimental temperature increases (ΔT) across substrates with different thermal conductivities, demonstrating good agreement between the two. ΔT is defined as the difference between the simulated temperature and room temperature. The simulated temperature was calculated as the average temperature at the top surface of the diamond, whereas the experimental temperature was obtained by averaging the measurements from five diamonds at 8 mW, as described in the experimental section. The substrate thermal conductivity values used in the figure were obtained from literature sources. The close match between simulated and experimental ΔT values confirms that the simulation model reliably reproduces the laser-induced heating observed experimentally.

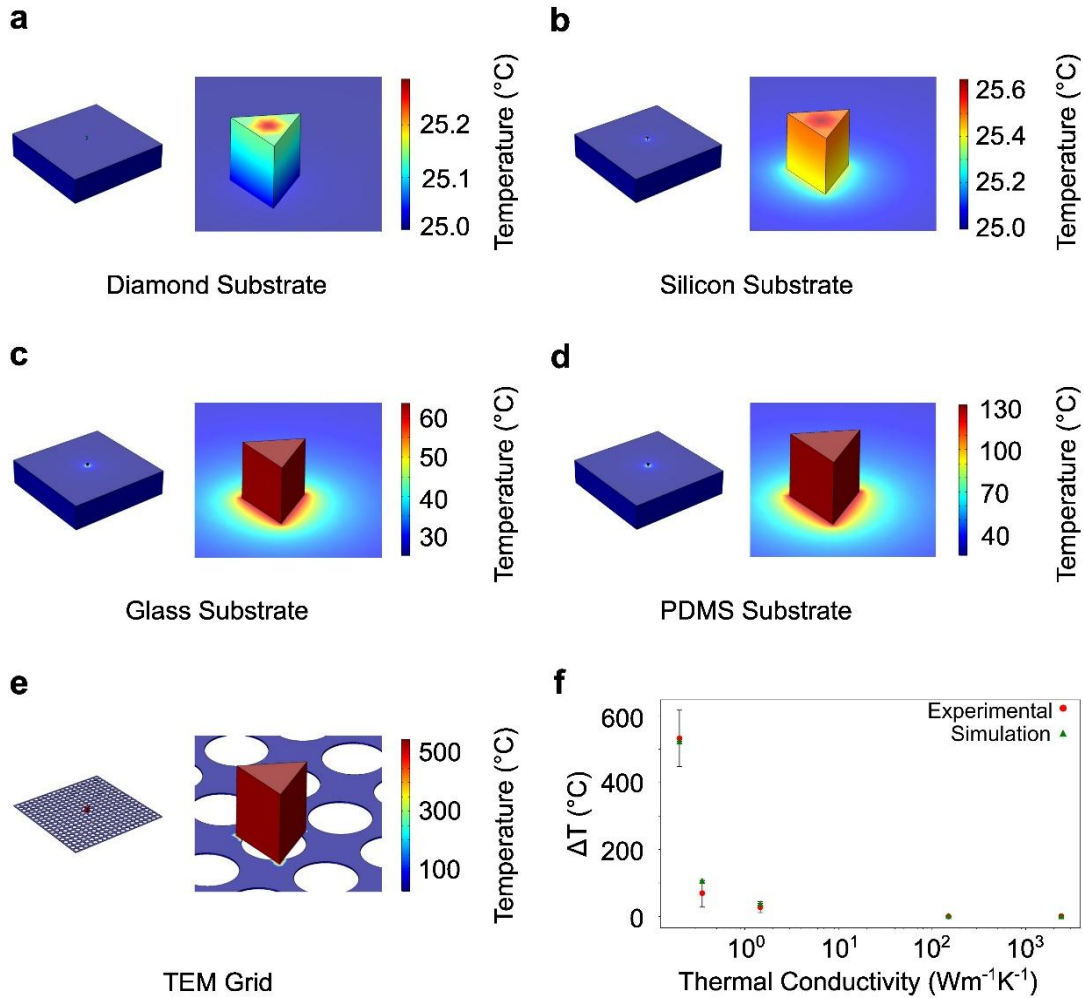

**Figure 4.** Simulated steady-state temperature distribution of a laser-heated SiV⁻ diamond on five different thermal conductivity substrates under 8 mW laser excitation using a 532 nm Gaussian beam. Panels **(a)**–**(e)** show temperature profiles for **(a)** bulk diamond, **(b)** silicon, **(c)** glass, **(d)** PDMS, and **(e)** thin amorphous holey carbon. Panel **(f)** compares the simulated and experimental temperature rise (ΔT) as a function of substrate thermal conductivity, where red markers represent experimental values and green markers represent simulation results. The error bars represent the standard deviation of the measurements taken at each point.

In the second part of the simulation, the temperature distribution of the diamond was analyzed on a silicon substrate with varying PMMA thicknesses under 8 mW laser excitation. The thermal conductivity, heat capacity, and density of PMMA were taken directly from the COMSOL materials library. **Figure 5 (a-e)** illustrates the simulation results for microdiamonds on a silicon substrate with different PMMA thicknesses, highlighting the influence of PMMA thickness on laser-induced heating of the microdiamonds. The simulation results show that increasing the PMMA thickness enhances laser-induced heating at the top surface of the diamond, due to reduced heat dissipation into the underlying silicon substrate, in agreement with the experimental observations discussed earlier. **Figure 5(f)** compares the simulated and experimental temperature increases (ΔT) for substrates with varying PMMA thicknesses. The close agreement between the two confirms that the simulation model accurately reproduces the experimentally observed laser-induced heating.

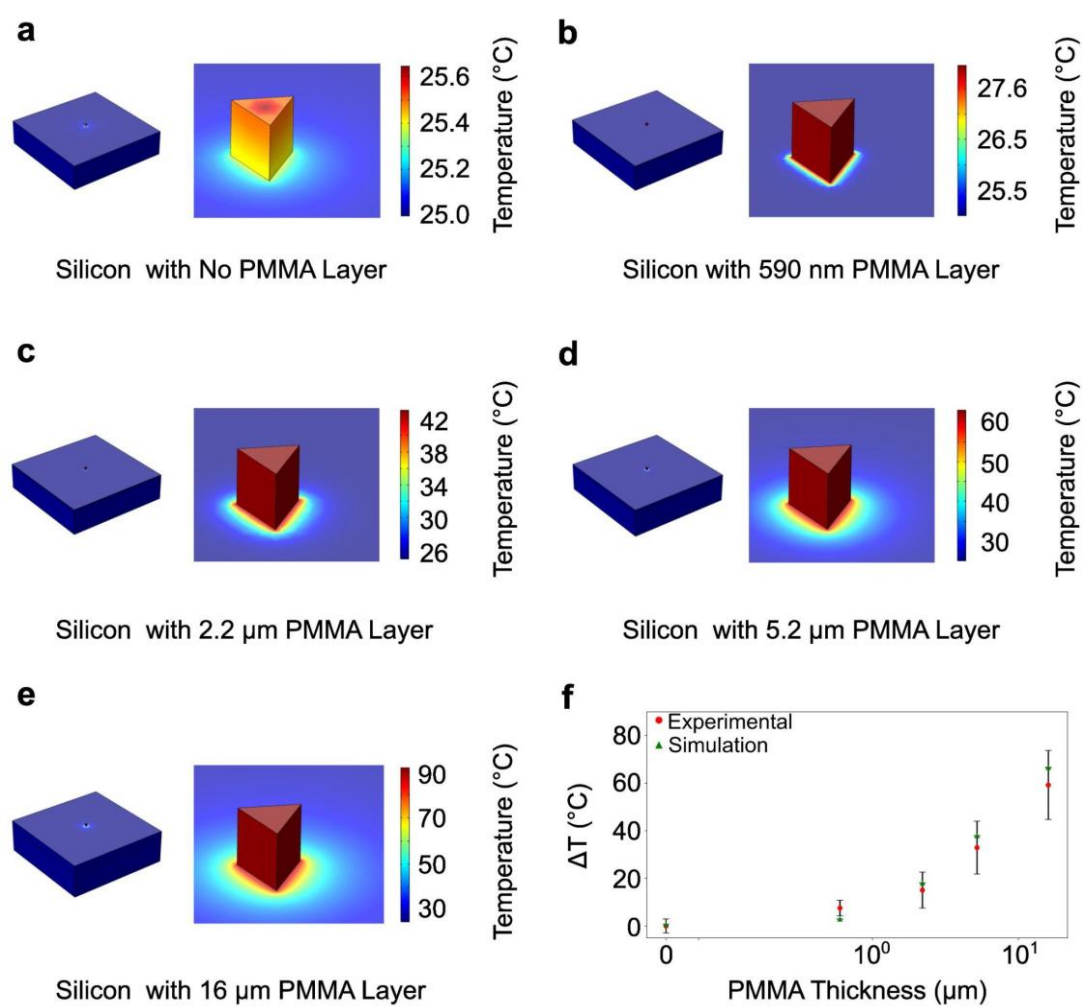

**Figure 5.** COMSOL-simulated steady-state temperature distributions of an SiV⁻ diamond under 8 mW, 532 nm laser excitation on substrates with varying PMMA residue thicknesses. Panels **(a)–(e)** display the thermal profiles for substrates with **(a)** no PMMA, **(b)** 590 nm PMMA, **(c)** 2.2 μm PMMA, **(d)** 5.2 μm PMMA, and **(e)** 16 μm PMMA. Panel **(f)** shows the comparison between experimental and simulated temperature rises (ΔT) as a function of effective thermal conductivity, where red points indicate experimental measurements and green points correspond to simulation results. The error bars represent the standard deviation of the measurements taken at each point.

Before concluding, it is essential to note that the thermal conductivity of the thin amorphous holey carbon was estimated from literature values for amorphous carbon thin films[38]. The actual thermal conductivity of the holey TEM grid used in the experiments could not be directly measured, which may introduce some uncertainty in the ΔT versus thermal conductivity plot as shown in **Figure 4(f).** Laser-induced heating on low-thermal-conductivity substrates is significant, leading to temperature rises of several hundred degrees within a very short time. To evaluate the timescale of this heating process, we recorded time-resolved PL spectra from a SiV$^-$ microdiamond placed on a thin amorphous holey carbon substrate under 1 mW excitation power, a regime where moderate heating had been observed in our previous experiments. Measurements taken at 10 ms intervals showed no observable shift in the ZPL, as further confirmed by 2D heat map analysis illustrated in **Figure S7 (cf. Supporting Information)**. This indicates that the effect of laser power on heating is instantaneous within our experimental resolution, and no waiting time is necessary when adjusting the excitation power. In the COMSOL simulation, to reduce computational load and improve visualization of the simulation results while maintaining physical accuracy, the substrate lateral dimensions and thickness were reduced to 200 μm and 50 μm, respectively, for bulk diamond, silicon, glass, and PDMS. For the thin amorphous holey carbon substrate, dimensions were set to 100 μm laterally and 100 nm in thickness. This reduction was verified not to influence the temperature profile.

## 3. Conclusions

In conclusion, we systematically investigated laser-induced optical heating in microdiamonds across five substrates with distinct thermal conductivities: bulk diamond, silicon, glass, PDMS, and thin amorphous holey carbon. With increasing laser power, the temperature rise in diamonds varies significantly depending on the thermal conductivity of the substrate. The lower the substrate thermal conductivity, the less laser power was required to observe local heating effects in the microdiamonds. It was found that glass and PDMS, with thermal conductivities of ~1.4 W m$^{-1}$ K$^{-1}$ and 0.35 W m$^{-1}$ K$^{-1}$ respectively, required approximately 2 mW of laser power to observe optical heating. In contrast, bulk diamond and silicon showed negligible heating even at 8 mW of laser power. Thin amorphous holey carbon, with the lowest thermal conductivity of 0.2 W m$^{-1}$ K$^{-1}$, exhibited significant laser-induced heating even with just 500 μW of power. In addition, the thickness of the substrate–sensor interfacial polymer layer was found to strongly influence optical heating, with even a 2.2 μm layer producing a significant temperature rise at 2 mW. The experimental results were further verified using COMSOL 3D heat transfer simulations. So, care should be taken when using high laser power on SiV$^-$ diamonds placed on low-thermal-conductivity substrates or in devices with residual polymer interlayers after nanofabrication, as localized heating can degrade the performance and reliability of optical thermal sensors introducing spurious effects on temperature measurements. These findings highlight the critical role of substrate selection and interfacial layers in controlling laser-induced heating of microdiamonds, providing practical guidance for optimizing thermal management in diamond-based thermometry, photonic, quantum, and sensing applications.

## 4. Methods

### 4.1. Sample Preparation

SiV⁻ diamonds used in this work were synthesized in C−H−Si (0.19 % atom) growth system using the high-pressure and high-temperature (HPHT) method described elsewhere.[40] Briefly, a powder mixture consisting of 300 mg of adamantane ($C_{10}H_{16}$, >99% purity, Sigma-Aldrich) and 18 mg of tetraphenylsilane ($C_{24}H_{20}Si$, 96% purity, Sigma-Aldrich) was manually ground for 5 minutes using a jasper mortar and pestle. Next, the mixture is compressed into a 65 mg pellet and placed inside a titanium capsule. A toroid-type high-pressure chamber is employed to generate the required high pressure and temperature (8 GPa/~2000 °C) in the reaction cell for 60s.[41] After the reaction, the sample is rapidly cooled to room temperature while maintaining the pressure. Finally, the microdiamonds were dispersed in isopropyl alcohol (IPA) for storage.

The prepared SiV diamonds were drop-cast onto various substrates with differing thermal conductivities—including bulk diamond, silicon, glass, PDMS Gel-Pak, and thin amorphous holey carbon film (TEM grid: TedPella_GYCU200) for optical characterization. To remove residual solvent and improve microdiamond–substrate adhesion, all substrates were heated on a hotplate at 200 °C for 20 minutes.

To prepare silicon substrates with various PMMA thicknesses, PMMA (A5, 950k) layers were deposited using dip coating techniques. PMMA films were deposited using a dip coater (Ossila L2006A2-US). Withdrawal speeds were adjusted to control the PMMA thickness: a speed of 0.5 mm/s produced the thickest coating of 16 μm, while 1 mm/s yielded 5.2 μm, 5 mm/s produced 2.2 μm, and 10 mm/s resulted in a 590 nm-thick layer. After coating, all samples were baked on a hotplate at 180 °C for 5 minutes to evaporate residual solvent and stabilize the PMMA layer. The thickness of the coated PMMA layer was measured using a profilometer (Dektak stylus), which provides high-resolution surface profiling through contact-based scanning.

## 4.2. Optical Measurement

Fluorescence and Raman signals from SiV diamonds were collected and analyzed using our custom optical setup, shown in **Figure 1(a)**. A 532 nm continuous-wave laser (Cobolt Samba) served as the excitation source. The laser beam was first guided from the source using a single-mode fiber (SM fiber) and two fiber couplers (FC), ensuring spatial mode quality and beam stability. To focus the laser spot onto the sample surface, a 4f system and a 100× objective (NA=0.7; Thorlabs, MY100X-806) were employed. The sample was placed on a three-dimensional micro-positioning stage. In this confocal setup, the laser beam was guided through some mirrors before being reflected by a dichroic mirror (Semrock FF649-Di01-25x36). The SiV emission collected from the sample by the same objective was transmitted through the dichroic mirror, directed by mirrors, and then split by a 50/50 beamsplitter between a spectrometer (Andor SR-500i) and an APD (Excelitas SPCM-AQRH-14-FC). Laser power was adjusted using a tunable neutral density (ND) filter placed in the excitation path. In the collection path, a notch filter (Semrock, NP-01-532RU-25) was placed to further filter out the 532 nm laser signal. The APD detected all incoming photons from the collection path, while a scanning mirror (Newport SFM-CD300B) was used to steer the laser beam across the sample. A confocal map of the scanning area was generated using LabVIEW software. The spectrometer recorded emission spectra from ensembles of emitters within the microdiamond, and the data were analyzed using Andor Solis software. In the first part of experiment, the laser power was gradually increased from 1 μW to 8 mW in fine steps (10 μW, 50 μW, 100 μW, 250 μW, 500 μW, 750 μW, 1 mW, 1.25 mW, 1.5 mW, 2 mW, 4 mW, and 6 mW) using a tunable neutral density (ND) filter. This allowed us to resolve changes in emission intensity across both the low- and high-power regimes. In the second part of the study, the laser power was varied over a reduced range (100 μW to 8 mW) with coarser increments (250 μW, 500 μW, 1 mW, 2 mW, and 4 mW), enabling faster acquisition while still capturing the key power-dependent behavior. The focused laser spot diameter was calculated by $d_{laser} = \frac{1.22 \times \lambda}{NA} = 0.93$ μm, where λ=532 nm is the laser wavelength and NA=0.7 is the numerical aperture of the objective used in our setup.

## 4.3. Comsol Simulation

Three-dimensional, steady-state heat transfer simulations were carried out using the Heat Transfer in Solids module of **COMSOL Multiphysics (v.5.5)** to model the temperature distribution in microdiamonds on different substrates under laser excitation. This module applies the finite element method (FEM) to numerically solve the stationary heat conduction equation, enabling spatially resolved temperature mapping under laser excitation. The SiV⁻ diamond was modeled with an octahedral geometry with an edge-to-edge dimension of 5 μm and was thermally coupled to the underlying substrate. In COMSOL, the general heat transport equation is given as

$$\rho \cdot C_p \cdot u \cdot \nabla T + \nabla \cdot q = Q \qquad (1)$$

where $\rho$ is the material density, $u$ is the velocity field for convective transport, $C_p$ is the specific heat capacity, $q$ is the heat flux ($q = -k \cdot \nabla T$), T is the temperature, and $Q$ is the heat source. For steady state analysis, temperature does not change with time meaning the transient term $C_p \cdot \frac{\delta T}{\delta t}$ is considered zero. Here, the laser served as a heat source which is material specific according the following equation[42]

$$Q(r) = \frac{2PA}{\pi w^2} \cdot e^{(-2r^2/w^2)} \qquad (2)$$

where *P* is the laser excitation power, *A* is the absorption coefficient of the microdiamond, *w* is the laser spot radius, and r is the radial distance from the beam center. The laser spot radius *w* was calculated using the diffraction-limited formula: $w = \frac{1.22 \times \lambda}{NA \times 2}$, where λ=532 nm is the laser wavelength and NA=0.7 is the numerical aperture of the objective used in our setup. The microdiamond absorptance *A* was estimated using the Beer–Lambert law: $A = 1 - e^{(-\alpha d)}$, where $\alpha = N_{siv} \times \sigma$ is the absorption coefficient, *d*=5 μm is the optical path length, $N_{siv}$ is the SiV center concentration, and $\sigma$ is the absorption cross-section at 532 nm. A value of $\sigma = 1 \times 10^{-16}$ was used for SiV centers under 532 nm excitation.[33] The SiV concentration was assumed to be $N_{siv} = 1.24 \times 10^{18}$ cm$^{-3}$, in close agreement with the experimentally synthesized concentration. For the simulation, the thermal conductivity of the microdiamond and substrates used in the experiments was obtained from the literature while the heat capacity and density of all materials—including the diamond, substrate, and any intermediate layers—were taken directly from the COMSOL material library. An *Extremely Fine* element size setting was applied to the diamond and its immediate contact region with the substrate to resolve fine temperature gradients induced by the laser, as seen from **Figure S7 (cf. Supporting Information)**. The initial temperature was set to 298.15 K, corresponding to room temperature.

**Data Availability Statement**

The datasets generated during and/or analyzed during the current study are available from the corresponding author on reasonable request.

**Funding:** T. T. T acknowledges the financial support from the Australian Research Council (DE220100487, DP240103127). T. T. T. and T. D. thank the Queensland Department of Environment, Science, and Innovation for their financial support (Q2032010).

**Author Contribution**

M.S.H. and T.T.T. conceived the project idea and designed the experiments. M.S.H., Y.C., and T.T.T. built the optical system and its software. E.E. fabricated SiV microdiamonds. M.S.H. performed the COMSOL simulation and analysis. M.S.H. and T.T.T. analyzed the data. T.T.T. supervised the project. All authors discussed the results and co-wrote the manuscript.

**Supporting Information**

**Laser-Induced Heating in Diamonds: Influence of Substrate Thermal Conductivity and Interfacial Polymer Layers**

*Md Shakhawath Hossain,[1] Jiatong Xu,[1] Thi Ngoc Anh Mai,[1] Nhat Minh Nguyen,[1] Trung Vuong Doan,[1] Chaohao Chen,[2] Qian Peter Su,[3] Yongliang Chen,[4] Evgeny Ekimov,[5] Toan Dinh,[6,7] Xiaoxue Xu,[3] and Toan Trong Tran [1, *]*

[1]School of Electrical and Data Engineering, University of Technology Sydney, Ultimo, NSW, 2007, Australia.

[2]School of Mathematical and Physical Sciences, Faculty of Science, University of Technology Sydney, NSW 2007, Australia

[3]School of Biomedical Engineering, University of Technology Sydney, Ultimo, NSW, 2007, Australia.

[4] Department of Physics, The University of Hong Kong, Pokfulam, Hong Kong, China.

[5]School of Engineering, University of Southern Queensland, Toowoomba, Queensland 4350, Australia

[6]Center for Future Materials, University of Southern Queensland, Toowoomba, Queensland 4350, Australia

*Corresponding author: trongtoan.tran@uts.edu.au

The Supporting Information includes:

**Supporting Information Figure S1:** Normalized Raman spectrum of a SiV⁻ diamond after annealing at 200 °C, fitted with a single Lorentzian profile.

**Supporting Information Figure S2:** Photoluminescence spectra and calibration curve of ZPL position as a function of temperature of a representative silicon-vacancy diamond on a bulk diamond substrate, measured at various temperatures.

**Supporting Information Figure S3:** Temperature versus laser power for five separate SiV⁻ diamonds, each measured on a different substrate.

**Supporting Information Figure S4:** Height profiles of PMMA films on silicon substrate

**Supporting Information Figure S5:** Temperature vs. laser power for five different SiV⁻ diamonds on silicon with varying PMMA thicknesses

**Supporting Information Figure S6:** Time resolved 2D heatmap of photoluminescence (PL) spectra from a single SiV⁻ diamond on thin amorphous holey carbon substrate under continuous 1 mW laser excitation.

**Supporting Information Figure S7:** Mesh design for the COMSOL simulation.

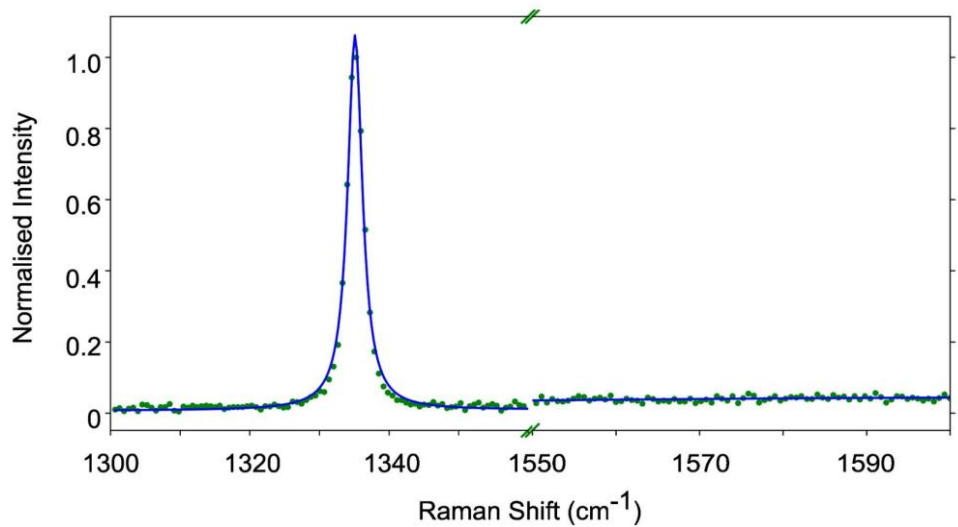

**Figure S1.** Normalized Raman spectrum of a SiV⁻ diamond after annealing at 200 °C, fitted with a single Lorentzian profile. The narrow, symmetric peak at ~1335 cm⁻¹ indicates high crystalline quality. The absence of a G-band near 1580 cm⁻¹ confirms the lack of graphitic or amorphous carbon. The spectrum was acquired with a 30 s integration time using 1.7 mW excitation power through the objective.

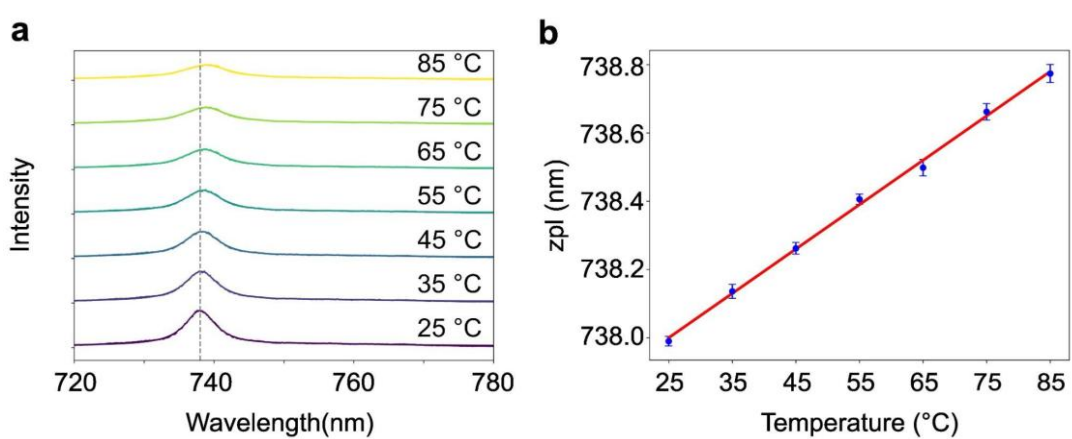

**Figure S2. (a)** Photoluminescence spectra of a representative silicon-vacancy diamond on a bulk diamond substrate, measured at various temperatures. Each spectrum is fitted with a double Lorentzian function for accurate peak analysis. Spectra were acquired with 1 s integration time and 500 µW excitation power. **(b)** Calibration curve of the zero-phonon line (ZPL) position as a function of temperature for diamonds on a bulk diamond substrate. Each data point represents the mean ZPL value obtained from Lorentzian fits to five diamonds at each temperature. Error bars denote the standard deviation of the ZPL positions. The red line indicates a linear fit used for temperature calibration.

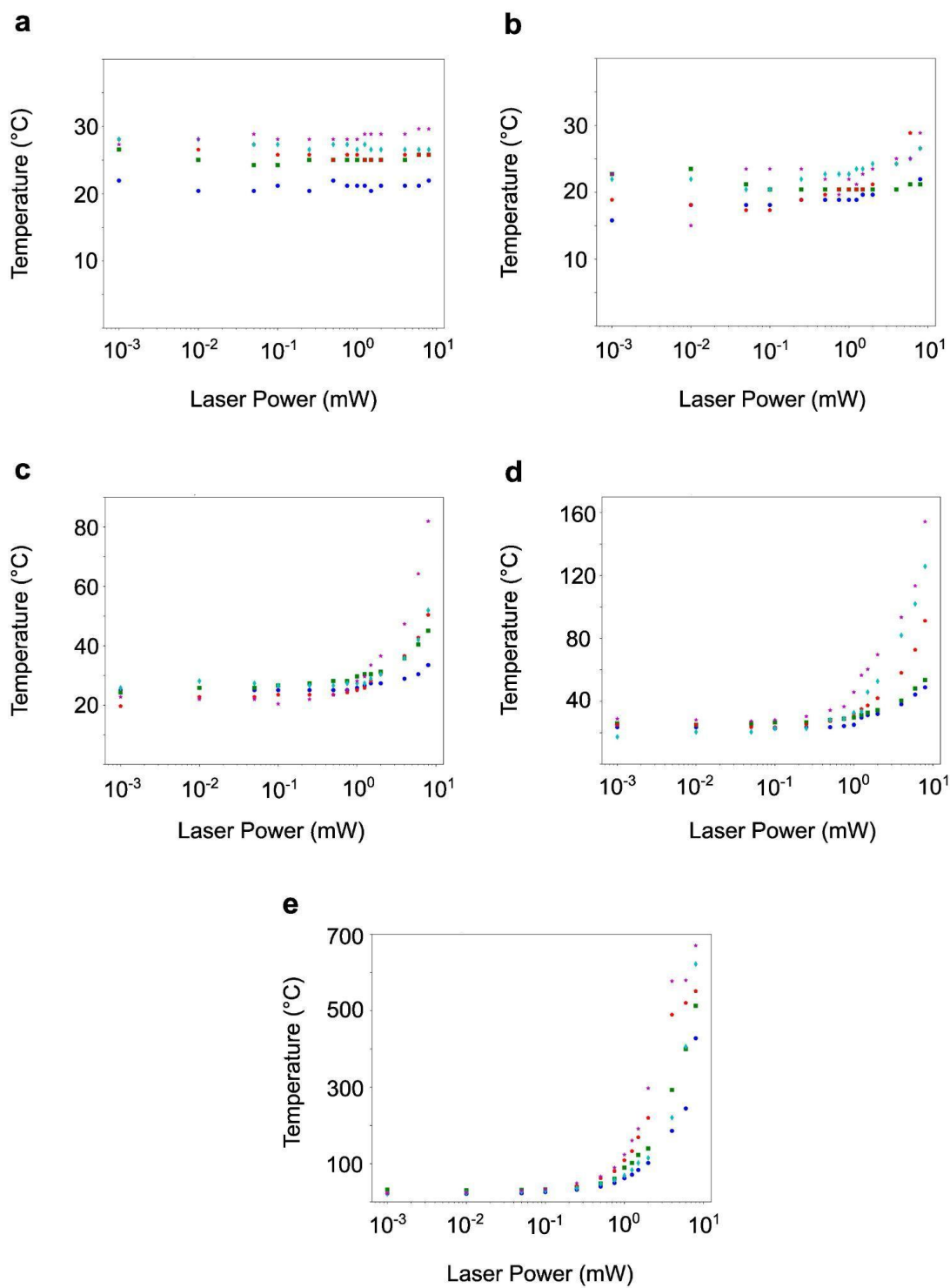

**Figure S3.** Temperature versus laser power for five separate SiV⁻ diamonds, each measured on a different substrate. Colors distinguish individual diamonds. Panels (a)–(e) correspond to bulk

diamond, silicon, glass, PDMS, and thin amorphous holey carbon, respectively. Laser power is shown on a logarithmic scale.

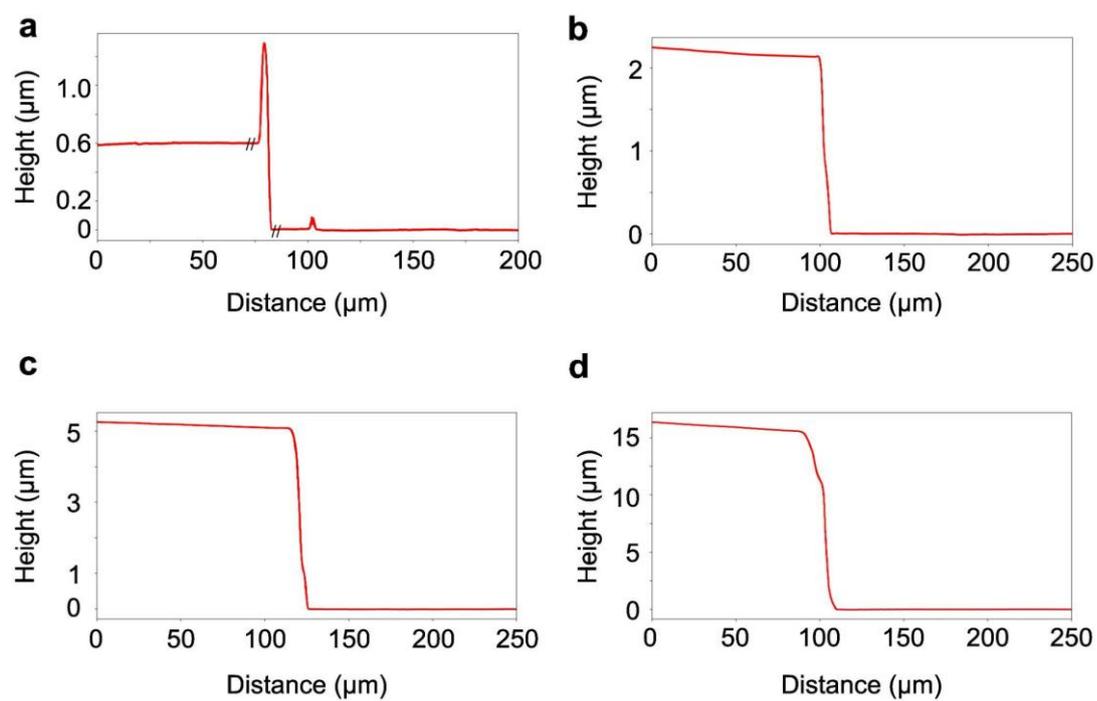

**Figure S4.** Height profiles of PMMA films on silicon substrate with thicknesses of approximately (a) 590 nm, (b) 2.2 μm, (c) 5.2 μm, and (d) 16 μm. Step heights were measured across a scratched edge using a stylus profilometer.

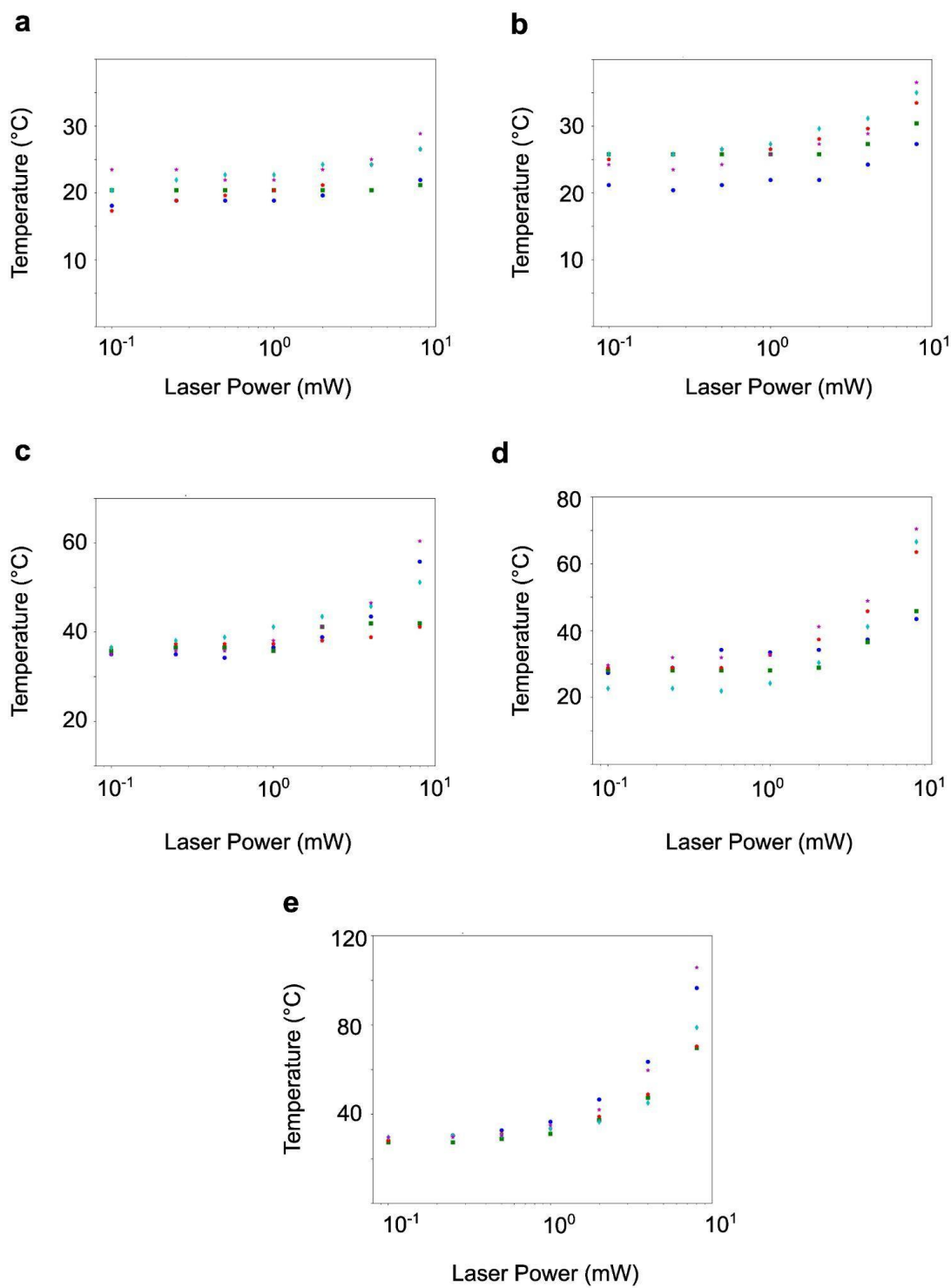

**Figure S5.** Temperature vs. laser power for five different SiV⁻ diamonds on silicon with varying PMMA thicknesses: (a) No PMMA, (b) 590 nm, (c) 2.2 µm, (d) 5.2 µm, (e) 16 µm. Colors denote individual diamonds. Laser power shown on a logarithmic scale.

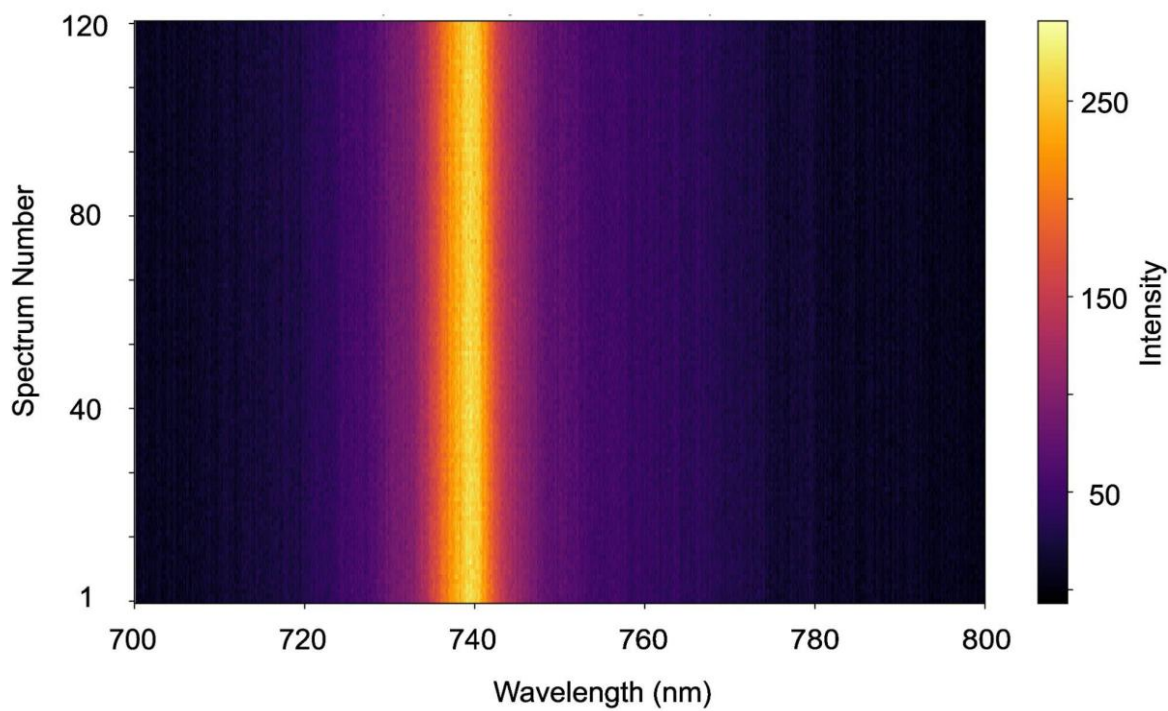

**Figure S6.** Time resolved 2D heatmap of photoluminescence (PL) spectra from a single SiV⁻ diamond on thin amorphous holey carbon substrate under continuous 1 mW laser excitation. A total of 120 spectra were recorded with 10 ms intervals (total acquisition time 1.2 s). The x-axis represents the wavelength range, while the y-axis corresponds to the spectrum index. The color scale indicates PL intensity.

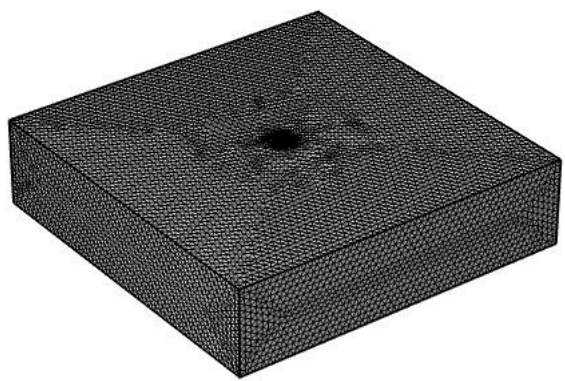

**Figure S7.** Mesh design for the COMSOL simulation. Tetrahedral mesh was used, with the smallest element size (40 nm) in the laser-illuminated region of the microdiamond and the largest element size (4 µm) in the bulk substrate.